\documentclass[cleveref]{lipics-v2021}

\newif\ifpreprint
\preprinttrue

\ifpreprint
  \hideLIPIcs
  \nolinenumbers
\fi

\usepackage{amsmath,amssymb,amsfonts}
\usepackage{mathtools}
\usepackage{graphicx}
\usepackage{textcomp}
\usepackage[dvipsnames,svgnames,x11names]{xcolor}
\usepackage{xspace}
\usepackage{color,enumerate,array,graphicx,tabularx,listings}
\usepackage{algorithm,algorithmicx}
\usepackage{rotating, multirow, makecell,adjustbox}
\usepackage{algpseudocode}
\usepackage[normalem]{ulem}
\usepackage{tikz}
\usepackage{dsfont}
\usepackage{forest}
\usepackage{multicol}
\usepackage{caption}
\usepackage{pgfplots}
\usepackage{pgfplotstable}
\pgfplotsset{compat=1.18}
\usepackage{float}
\usepackage{cuted}
\usepackage[most]{tcolorbox} 

\usepackage[T1]{fontenc}
\usepackage{lmodern}

\newtheorem{problem}{Problem}

\tcbuselibrary{theorems}

\tcolorboxenvironment{problem}{
  enhanced,
  breakable,
  colback=black!3,     
  colframe=black!50,   
  boxrule=0.8pt,
  arc=2pt,
  left=0.8em,
  right=0.8em,
  top=0.6em,
  bottom=0.6em,
  before skip=0.8\baselineskip,
  after skip=0.8\baselineskip,
}

\newcommand{\builddag}[1]{\textsc{BuildDAG}({#1})\xspace}
\newcommand{\preprocess}[1]{\textsc{ScoringOBS}({#1})\xspace}
\newcommand{\findschedulePone}[1]{\textsc{ScheduleOBS}({#1})\xspace}
\newcommand{\findschedulePtwo}[1]{\textsc{SchedulePBC}({#1})\xspace}
\newcommand{\scoring}[1]{\textsc{ScoringPBC}({#1})\xspace}

\newcommand{\gasused}[1]{\ensuremath{{#1}.\mathit{gasU}}\xspace}

\newcommand{\tip}[1]{\ensuremath{{#1}.\mathit{tip}}\xspace}

\newcommand{\mempool}{\ensuremath{\mathit{Mpool}}\xspace}
\newcommand{\makespan}[1]{\ensuremath{{#1}.\mathit{mspan}}\xspace}
\newcommand{\reward}[1]{\ensuremath{{#1}.\mathit{reward}}\xspace}
\newcommand{\readset}[1]{\ensuremath{{#1}.\mathit{R}}\xspace}
\newcommand{\writeset}[1]{\ensuremath{{#1}.\mathit{W}}\xspace}
\newcommand{\executiontime}[1]{\ensuremath{{#1}.\mathit{t}}\xspace}
\newcommand{\id}[1]{\ensuremath{{#1}.\mathit{id}}\xspace}

\newcommand{\indegree}[1][]{\ensuremath{\mathit{indg}{#1}}\xspace}
\newcommand{\outdegree}[1][]{\ensuremath{\mathit{outdg}{#1}}\xspace}
\newcommand{\degree}[1][]{\ensuremath{\mathit{d}{#1}}\xspace}
\newcommand{\successor}[1][]{\ensuremath{\mathit{succ}{#1}}\xspace}
\newcommand{\predecessor}[1][]{\ensuremath{\mathit{pred}{#1}}\xspace}
\newcommand{\height}[1][]{\ensuremath{\mathit{hgt}{#1}}\xspace}
\newcommand{\volume}[1][]{\ensuremath{\mathit{vol}{#1}}\xspace}
\newcommand{\priority}[1][]{\ensuremath{\mathit{pri}{#1}}\xspace}
\newcommand{\start}[1][]{\ensuremath{\mathit{start}{#1}}\xspace}
\newcommand{\finish}[1][]{\ensuremath{\mathit{end}{#1}}\xspace}
\newcommand{\core}[1][]{\ensuremath{\mathit{core}{#1}}\xspace}
\newcommand{\conflicts}[1][]{\ensuremath{\mathit{confs}{#1}}\xspace}
\newcommand{\deferred}{\ensuremath{\mathit{deferd}}\xspace}
\newcommand{\selected}{\ensuremath{\mathit{sel}}\xspace}

\newif\ifannote
\annotetrue  

\ifannote
    
    \newcommand{\anncomment}[3]{{\color{#1}[#2: #3]}}
\else
    
    \newcommand{\anncomment}[3]{}
\fi

\title{Exploiting Multi-Core Parallelism in Blockchain Validation and Construction}

\author{Arivarasan Karmegam}{IMDEA Networks Institute, Madrid, Spain \and Universidad Carlos III de Madrid, Madrid, Spain}{arivarasan.karmegam@networks.imdea.org}{https://orcid.org/0000-0002-6690-0285}{}
\author{Lucianna Kiffer}{IMDEA Networks Institute, Madrid, Spain}{}{https://orcid.org/0000-0003-2022-7993}{}
\author{Antonio {Fernández Anta}}{IMDEA Software Institute, Madrid, Spain \and IMDEA Networks Institute, Madrid, Spain}{}{https://orcid.org/0000-0001-6501-2377}{}
\authorrunning{Karmegam et al.}

\Copyright{Authors}


\acknowledgements{We would like to thank Lioba Heimbach and the DISCO group at ETH Zurich for providing the transaction traces used in our experiments section.
This work is funded in part by MICIU/AEI /10.13039/501100011033/, ERDF, and the ESF+ under predoctoral training grant PREP2022-000373 and grant DRONAC (PID2022-140560OB-I00). 
Grant CEX2024-001471-M funded by MICIU/AEI/10.13039/501100011033.}


\begin{document}

\maketitle

\begin{abstract}
Blockchain validators can reduce block processing time by exploiting multi-core CPUs, but deterministic execution must preserve a given total order while respecting transaction conflicts and per-block runtime limits. This paper systematically examines how validators can exploit multi-core parallelism during both block construction and execution without violating blockchain semantics.  
We formalize two validator-side optimization problems: (i) executing an already ordered block on \(p\) cores to minimize makespan while ensuring equivalence to sequential execution; and (ii) selecting and scheduling a subset of mempool transactions under a runtime limit \(B\) to maximize validator reward. For both, we develop exact Mixed-Integer Linear Programming (MILP) formulations that capture conflict, order, and capacity constraints, and propose fast deterministic heuristics that scale to realistic workloads. 

Using Ethereum mainnet traces and including a Solana-inspired declared-access baseline (Sol) for ordered-block scheduling and a simple reward-greedy baseline (RG) for block construction, we empirically quantify the trade-offs between optimality and runtime. MILPs quickly become intractable as heterogeneity or core count increases, whereas our heuristics run in milliseconds and achieve near-optimal quality. For ordered-block execution, heuristic makespans are typically within a few percent of the MILP solutions (and can even surpass the MILP incumbent when the solver times out), yielding up to $1.5$ speedup with $p=2$ and $2.3$ speedup with $p=8$ over sequential execution, despite tight ordering constraints. For block construction, the heuristic achieves $99$--$100\%$ of the MILP optimum reward on homogeneous workloads, and $74$--$100\%$ of an LP-relaxation upper bound on heterogeneous workloads, where exact optimization often times out. The resulting block-construction throughput scales close to linearly with $p$, reaching up to $7.9$ speedup with $p=8$ in our experiments. These results demonstrate that lightweight, conflict-aware scheduling and selection can unlock substantial parallelism in blockchain validation, bridging the gap between sequential execution and the true potential of multi-core hardware.
\end{abstract}


\newpage
\section{Introduction} \label{sec:introduction}

\subsection{Motivation}

Blockchains process a continuous stream of user transactions that modify a shared replicated state. In most deployed designs (e.g., Ethereum-style execution), validators\footnote{Nodes responsible for proposing or verifying blocks, and executing transactions.} validate and execute blocks sequentially following a \emph{fixed total order.}
While this simplifies reasoning about state, it underutilizes modern multi-core CPUs. In practice, many transactions affect disjoint sets of accounts or storage slots and can be safely executed in parallel without altering the final state. Sequential execution, therefore, limits throughput and increases time-to-finality (i.e., the delay from block proposal to having its transactions finalized on-chain).
However, parallelizing block execution is not trivial, since the blockchain state is shared. Reordering or executing concurrently two transactions that conflict\footnote{Both write, or one writes and the other reads, the same account or storage slot.} 
when accessing the state may change the final state. Any viable solution must therefore preserve deterministic, order-equivalent semantics while exploiting conflict-free concurrency.

There is also an economic dimension when validators build blocks. They select transactions from a \emph{mempool} of pending transactions with heterogeneous runtimes and rewards. Since blocks
are limited in size by execution time (e.g., with a \textit{gas budget} or \textit{runtime limit}),
the question for validators is not ``how fast can we execute a given block?'' but ``how do we select the transactions in a block to maximize the reward?''

In this paper, we study how a validator can leverage multi-core hardware without violating blockchain semantics or economic constraints. We study two complementary validator-side optimization problems, one with a fixed block order and one with flexible selection:

\smallskip
\noindent\textbf{Ordered-Block Scheduling (OBS). 
}
Given a block with a fixed transaction set and total order, how should a validator exploit its cores to minimize total execution time while guaranteeing equivalence to sequential execution?
This is a \textbf{scheduling} problem: transactions that access disjoint sets of accounts or storage slots can run together, while conflicting transactions must respect the block order. We capture conflicts through a \textbf{dependency DAG} and aim for a parallel schedule that minimizes the time to execute the whole block (\emph{makespan}), leveraging the inherent parallelism present in existing blocks.
This is a classical scheduling problem known as Precedence-Constrained Scheduling, known to be NP-hard on 3 cores and on 2 cores when transactions have execution times in \{1,2\}\cite{garey2002computers}.

\smallskip
\noindent\textbf{Parallel-Block Construction (PBC). 
}
On the other hand, when assembling a new block under a runtime limit, the validator has to choose from a mempool of pending heterogeneous transactions that differ in both reward and execution time. Here, the problem is deciding which transactions should be included, and how should they be assigned to cores, so that 
(i) conflicts are respected, (ii) the overall schedule fits within the runtime limit, and (iii) the total validator reward is maximized. Unlike with OBS, we can now trade off which transactions to include against how efficiently they can be scheduled in parallel. This couples selection and scheduling, reflecting the real economic and system constraints that a validator faces in practice.
To our knowledge, prior work has not studied this specific combination of conflict constraints, runtime limit, and validator reward maximization. Observe that PBS is NP-hard, since it generalizes classical NP-hard problems such as knapsack with conflict graphs, and parallel-machine scheduling with conflicts.

\subsection{Contributions}

    \noindent
    \textbf{(1) Problem definitions:} We \textbf{formalize two validator-side optimization problems:} (i)~Ordered-Block Scheduling (OBS), scheduling an ordered block on $p$ cores minimizing makespan, and 
    (ii)~Parallel-Block Construction (PBC), selecting and scheduling a subset of mempool transactions on $p$ cores under a runtime limit $B$ to maximize validator reward, all while respecting read/write conflicts.

    \noindent
    \textbf{(2) MILP formulations:} We present exact \textbf{Mixed-Integer Linear Programming (MILP) formulations}\footnote{Some formulations use only integer or Boolean variables. We refer to all of them as MILP for simplicity.} that capture conflicts and precedence, core capacity, and budget constraints. These MILPs provide optimal baselines on small-to-medium instances, but become intractable at scale; we therefore use them primarily for benchmarking and insight.

    \noindent
    \textbf{(3) Fast heuristics:} We design \textbf{deterministic heuristics} that produce feasible schedules rapidly, trading optimality for orders-of-magnitude higher speed.
    E.g., the heuristic for OBS is guaranteed to find a
schedule with makespan at most $2-1/p$ times the optimal.

    \noindent
    \textbf{(4) Empirical study:} 
    We quantify empirically
the limits of parallelism in block execution and construction.
Using Ethereum mainnet traces, we compare MILPs and heuristics on both the time taken to find a
solution and its quality.
We include in the comparison a Solana-inspired declared-access baseline (Sol) for OBS and a simple reward-greedy baseline (RG) for PCB, instantiated using the access sets extracted from Ethereum traces.

We observe that using multiple cores can significantly reduce makespan and increase the block reward relative to sequential block execution.
For instance, for OBS we observe that with only 2 cores we can reduce the makespan by a factor of $1.57$, and with 8 cores by more than $2$.
For PBC, the throughput can be increased almost linearly with the number of cores, and the reward
increased by a factor of more than $2$ even with $4$ cores.
We also observe that the runtime for solving the MILPs grows steeply with transaction heterogeneity, core count, and runtime limit. 
However, our heuristics achieve makespans and rewards close to MILP optima when available (and remain competitive against LP-relaxation upper bounds when exact optimization times out), while running in milliseconds.
    





\subsection{Structure of the Rest of the Paper}
The article is structured as follows: 
Section~\ref{sec:related} reviews existing literature related to this work.
Section~\ref{sec:problemstatement} presents in detail our model and problems statements.
Section~\ref{sec:problem1} 
discusses the heuristics for OBS.
Section~\ref{sec:problem2} presents the MILP formalization of PBC and discusses the heuristics for this problem.
Section~\ref{sec:experiments} describes the experiment details.
Finally, Section~\ref{sec:results}, presents and discusses the experimental 
results, while Section~\ref{sec:conclusion} concludes the paper.
%

\section{Related Work}
\label{sec:related}

Parallel execution of transactional workloads under determinism and conflict constraints naturally gives rise to classical combinatorial optimization problems in multiprocessor scheduling and packing. In this work, blockchain execution serves as a motivating application domain in which such constraints arise explicitly and at scale. Similar constraints for replica determinism have long been studied in dependable and replicated systems, where concurrent workloads must produce identical outcomes across replicas to ensure state consistency~\cite{jimenez2000deterministic_scheduling}.

At its core, OBS is a precedence-constrained makespan scheduling problem on identical processors $\left(P|prec|C_{max}\right)$. 
Multiprocessor scheduling with precedence and resource constraints is NP-hard in general~\cite{ullman1973NPcompletescheduling,GareyandJohnson1975complexity_results}, 
which explains why exact optimization becomes intractable for realistic block sizes and motivates heuristic or approximate approaches. A standard scalable approach is list scheduling, 
with $(2-1/p)$ approximation ratio~\cite{graham1966bounds,graham1969bounds}. 
Although this bound applies without additional resource conflicts, it motivates greedy event-driven schedulers as practical baselines when optimal solutions are infeasible. 
More broadly, the $P|prec|C_{max}$ literature shows that the structure of the precedence partial order 
can significantly affect complexity and approximability~\cite{prot2018survey_prec_structures,pinedo2016scheduling}. 
Our precedence DAGs arise from intersecting a total order with a symmetric conflict relation and do not align with known polynomial-time special cases, inheriting the general NP-hardness.
Our goal is thus to understand how simple, fast heuristics behave on realistic conflict structures and workload heterogeneity.

A defining constraint in blockchain execution is fixed-order determinism: correctness requires equivalence to a specific total order, not merely serializability to some order. Similar constraints arise in parallel state machine replication and deterministic transactional systems, where concurrency must respect an externally imposed total order~\cite{kotlaandDahlin2004highthroughputBFT,abadi2014deterministicdatabasesystems, Thomson2010determinism}. This motivates modeling concurrency as a precedence-constrained scheduling problem rather than allowing arbitrary reordering.

Several blockchain systems exploit parallelism to accelerate execution of an already ordered block. Speculative frameworks such as Block-STM~\cite{BlockSTM}, used in Diem~\cite{diemcodebase} and Aptos~\cite{aptoscodebase,aptoswhitepaper}, execute transactions optimistically in parallel and resolve conflicts via deterministic abort-and-retry against a preset transaction sequence. Related approaches such as BTM~\cite{anjana_AFT_Effecient_parallel_execution} use read/write-set hints to reduce aborts when access sets are known in advance. Other designs rely on explicit conflict declarations: Solana requires transactions to declare their access list, enabling the runtime (Sealevel~\cite{solanaSealevel}) to schedule non-overlapping transactions in parallel while serializing conflicts~\cite{yakovenko2017solana}. Object-centric execution follows a similar philosophy, representing state as objects and tracking object-level conflicts; Sui Lutris, in particular, exploits object-level access metadata to enable parallelism while resorting to consensus only when shared-object conflicts require total ordering~\cite{sui}. Across these designs (including recent systems such as Monad~\cite{monadParallelExecutionDocs2025}) parallelism is exploited opportunistically, but remains constrained by the prefix of the total order: a conflicting early transaction can stall later independent ones rather than being bypassed. OBS makes this restriction explicit by allowing any order-equivalent execution schedule, thereby formalizing the gap between opportunistic in-order parallelism and optimal makespan-minimizing execution under limited cores.

A substantial body of work studies scheduling under conflict graphs. For unit-time jobs, conflict-graph scheduling reduces to graph coloring, while heterogeneous execution times break this equivalence~\cite{batchscheduleexecute}. Empirical studies highlight sensitivity to contention and workload structure~\cite{sathyaperi2025graphcolouring}, and measurements on Ethereum show conflict graphs dominated by hotspots, limiting achievable speedups~\cite{anjana_AFT_Effecient_parallel_execution,heimbach2023definftshinderblockchain,Biton2025conflictsgraphedethereum}. Complementary semantic analyses characterize when reordering is safe under read/write effects, abstracting away from optimization objectives such as makespan or resource-aware selection~\cite{bartoletti2020concurrentsmartcontractexecution,bartoletti2021transactionparallelisminblockchains}. Closely related classical models consider identical-machine scheduling with incompatibility graphs $P|G|C_{max}$, which is NP-hard in general~\cite{blazewicz1991schedulingincompatiblejobs,kowalczyk2017exact_parallel_conflicts,brauner2016scheduling_conflictgraph}, with exact and MILP-based methods applicable only at moderate scales~\cite{moura2025compact_conflicts}.


PBC builds on these models by jointly selecting and scheduling transactions under a hard runtime budget. It generalizes multiple knapsack~\cite{chekuri2005polynomial}, knapsack with conflict graphs~\cite{pferschy2009knapsack}, and parallel-machine scheduling with rejection~\cite{engels2003techniques_scheduling_rejection,jansen2010scheduling_rejection_parallel}. Unlike existing block construction pipelines, which are typically driven by greedy fee-based selection (e.g., gas price or fee per gas) and largely ignore parallelizability and interaction effects among transactions~\cite{neiheiser2025anthemiusefficientmodular}, our formulation explicitly captures the interaction between transaction selection, conflicts, heterogeneous execution times, and core-level makespan. To our knowledge, deployed block assembly pipelines do not incorporate parallelizability directly into transaction selection in a systematic way, making block construction the primary systems gap our work targets.

\section{System Model and Problems Definitions} \label{sec:problemstatement}

\subsection{Blockchains}

We model a blockchain as a distributed ledger that maintains a shared state and processes a continuous stream of client-submitted transactions. Pending transactions reside in a \textit{mempool} until they are selected for inclusion in a block. Each block is proposed by a designated \textit{validator} (or leader) that assembles transactions from the mempool, subject to 
an upper bound on block execution time.



\subsection{Transactions}


We use the transaction model of Ethereum.
Each transaction $\tau$ has execution and fee parameters that govern how it is processed by the validator. 
When executed, $\tau$ consumes a certain amount of gas $\gasused{\tau}$. It also has a \emph{Priority Fee}, denoted $\tip{\tau}$, so that $\reward{\tau} = \gasused{\tau} \cdot \tip{\tau}$ is the reward to the validator that includes $\tau$ in a block.


\smallskip
\noindent We consider two execution-time regimes:
\begin{itemize}
\item  \textbf{Homogeneous} transactions with identical execution time (equivalently identical \gasused{\tau}).
\item  \textbf{Heterogeneous} transactions with varying execution times (equivalently varying \gasused{\tau}).
\end{itemize}

In Ethereum-like settings, simple transfers are often homogeneous, whereas smart contract transactions are typically heterogeneous.
We assume that when a transaction $\tau$ is issued, the client includes an \emph{access list} 
specifying the set of state keys (accounts and storage slots) $\readset{\tau}$ and $\writeset{\tau}$ that $\tau$ reads and writes, respectively. In practice, transfers typically access only a few keys (e.g., sender and receiver of cryptocurrency), whereas smart contract transactions may access many keys. 


\textbf{Conflicts.}
We say that two transactions $\tau_i$ and $\tau_j$ conflict if and only if they access a common state key and at least one of them writes to it. I.e., read--read overlaps are not conflicts, but any overlap involving a write is. Formally,
\begin{align}
    \mathrm{conflict}(\tau_i,\tau_j) \equiv \Big( (\writeset{\tau_i}\cap \writeset{\tau_j}) \cup (\writeset{\tau_i}\cap \readset{\tau_j}) \cup (\readset{\tau_i}\cap\writeset{\tau_j}) \Big) \neq\varnothing. & \label{subsubsec:conflicts}
\end{align}
%


\subsection{Problems Definitions}
Our goal is to exploit multi-core hardware to validate and execute transactions efficiently while preserving blockchain semantics.
%
We assume a validator has an execution environment that can process up to $p$ transactions in parallel, where $p$ is the number of CPU cores of the validator.
We also assume that the time to execute a transaction $\tau$ is proportional to its gas consumption, which we model as exactly $\gasused{\tau}$.
Then, we can define:

\begin{definition}[Schedule]
A schedule $S$ of a set of transactions $T$ on $p$ cores assigns to each transaction in $T$ a core and a starting time to run.  
    Given a schedule $S$ of a set of transactions $T$ on $p$ cores, its \emph{makespan}, denoted $\makespan{S}$, is the time to complete the execution (without preemption) of all transactions in $T$ as defined by the schedule $S$.
\end{definition}

We now formalize the two problems.

\begin{problem}
\label{prb:block}
\textbf{Ordered-Block Scheduling (OBS):} Given a block with the sequence of transactions $T=(\tau_1, \ldots, \tau_n)$, find a schedule $S$ of the transactions in $T$ on $p$ cores, that minimizes makespan $\makespan{S}$, subject to the constraint that conflicting transactions are not executed in parallel, and they are executed in the order of the sequence. 
\end{problem}

\begin{problem}
\label{prb:mpool}
\textbf{Parallel-Block Construction (PBC):} Given a runtime limit $B$ and a set of transactions in the mempool $\mempool=\{\tau_1, \ldots, \tau_n
\}$, select a subset of transactions $T \subseteq \mempool$ and a schedule $S$ of $T$ on $p$ cores such that: conflicting transactions are not executed in parallel,  $\makespan{S} \leq B$, and the total reward $\reward{T}=\sum_{\tau \in T} \reward{\tau}$ is maximized.
\end{problem}





\section{Ordered-Block Scheduling (OBS)} \label{sec:problem1}

In this section we propose algorithms to solve the Ordered-Block Scheduling (OBS) problem. All solutions start by creating a dependency graph. Given a block containing an ordered sequence of transactions $T=(\tau_1, \ldots, \tau_n)$, we construct a \textbf{directed dependency graph} $G=(T,E)$, where each node represents a transaction and each directed edge $(i,j) \in E$ indicates that transaction $\tau_i$ must precede $\tau_j$. An edge exists if (i) the two transactions conflict (as defined in \Cref{subsubsec:conflicts}), and (ii) $i<j$, ensuring consistency with the block’s total order. This graph captures all read/write dependencies that restrict concurrent execution. 
 

\subsection{OBS MILP Formulations}
\label{sec:OBS_MILP}

\begin{figure}[t!]
\centering
\begin{tcolorbox}[enhanced, arc=1mm, boxrule =0.8pt, colback=white, colframe=black, width=\textwidth, left =1pt, right=6pt,  ams align,fontupper=\footnotesize,]
\min \quad & \sum_{r \in [R]} y_{r} 
    && && \text{(Minimize the number of rounds used)} \label{obj:assign} \\[0.5ex]
\text{s.t.}\quad 
& \sum_{r \in [R]} x_{ir} = 1 
    && \forall i \in [n] 
    && \text{(Schedule each transaction once)} \label{con:onetrans} \\
&\sum_{i\in[n]} x_{ir} \leq p \cdot y_r 
    && \forall r \in [R]
    && \text{(At most $p$ transactions per $r$ used)} \label{con:p-cores} \\
& \sum_{t=1}^r x_{jt} - \sum_{t=1}^{r-1} x_{it} \leq 0
    && \forall (i,j) \in E, \forall r \in [R]
    && \text{($j$ cannot be scheduled before $i$)} \label{con:order} \\
& y_r \geq y_{r+1} &&  \forall r \in [1 .. R-1 ]
    && \text{(Consecutive round usage)} \label{con:no-gaps}\\
& x_{ir},y_r \in \{0,1\} 
    && \forall i \in [n],\; \forall r \in [R] 
    && \text{(Binary assignment)} \label{con:binary-assign} 
\end{tcolorbox}
\captionsetup{labelformat=simple,labelsep=colon,name={MILP}, justification=centering}
\caption{Ordered-Block Scheduling - Homogeneous Transactions}
\label{milp-pbm1-simple}
\end{figure}

\begin{figure}[t!]
\centering
\begin{tcolorbox}[enhanced, arc=1mm, boxrule =0.8pt, colback=white, colframe=black, width=\textwidth, left =1pt, right=6pt,  ams align,fontupper=\footnotesize,]
\min \quad & M 
    && && \text{(Minimize makespan)} \label{obj:min-mspan-c1} \\[0.5ex]
\text{s.t.}\quad
& \sum_{c \in [p]} x_{ic} = 1 
    && \forall i \in [n] 
    && \text{(Each trans.~in 1 core)} \label{con:onetrans-c1} \\
& s_i \geq 0 
    && \forall i \in [n] 
    && \text{(Non-negative start)} \label{con:noneg-c1} \\
& s_j \geq e_i 
    && \forall (i,j) \in E 
    && \text{(Precedence constr.)} \label{con:order-c1} \\
& e_i = s_i + t_i 
    && \forall i \in [n] 
    && \text{(End time Definition)} \label{con:runtime-c1} \\
& M \geq e_i 
    && \forall i \in [n] 
    && \text{(Makespan is max end)} 
    \label{con:mspan-c1} \\
&w_{ijc} \leq x_{ic} 
    &&  \forall (i,j) \in \mathcal{P}_{\text{non}} , \forall c \in [p]\\
&w_{ijc} \leq x_{jc} 
    &&  \forall (i,j) \in \mathcal{P}_{\text{non}} , \forall c \in [p]\\
&w_{ijc} \geq x_{ic} + x_{jc} -1 
    &&  \forall (i,j) \in \mathcal{P}_{\text{non}}, \forall c \in [p]\\
&z_{ij} =\sum_{c \in [p]} w_{ijc} 
    &&  \forall (i,j) \in \mathcal{P}_{\text{non}}
    && \text{(Same core flag)}\\
&e_i \le s_j + B \bigl( (1 - y_{ij}) + (1-z_{ij}) \bigr), 
    && \forall (i,j) \in \mathcal{P}_{\text{non}}\label{con:order1}\\
&e_j \le s_i + B \bigl( y_{ij} + (1-z_{ij}) \bigr), 
    && \forall (i,j) \in \mathcal{P}_{\text{non}}\label{con:order2}  \\
& M,\, s_i,\, e_i \in \mathbb{N} 
    && \forall i \in [n] 
    && \text{(Continuous variables)}  \label{con:linear-assign-c1} \\
& x_{ic} \in \{0,1\} 
    && \forall i \in [n],\; \forall c \in [p] 
    && \text{(Binary assignment)} \label{con:binary-assign-c1} \\
&w_{ijc} \in \{0,1\} 
    && \forall (i,j) \in \mathcal{P}_{\text{non}}, \forall c \in [p]\\
&y_{ij},z_{ij} \in \{0,1\} 
    && \forall (i,j) \in \mathcal{P}_{\text{non}}
\end{tcolorbox}
\captionsetup{labelformat=simple,labelsep=colon,name={MILP}}
\caption{Ordered-Block Scheduling - Homogeneous Transactions. Let $\mathcal{P}_{\text{non}}:=\{ (i,j) : i<j \land (i,j) \notin E \land (j,i) \notin E \}$ be the set of pairs of transactions without conflicts.}
\label{milp-pbm1-simple&complex}
\end{figure}

\subsubsection{Homogeneous Transactions}

We first consider the homogeneous case in which all the transactions have identical execution time. Since each transaction takes the same amount of time $\executiontime{\tau}$, we discretize time into rounds $R$ of uniform length. MILP \ref{milp-pbm1-simple} defines the corresponding MILP formulation. Here, $R$ denotes the maximum number of rounds, which must be an upper bound of the rounds required and has to be provided. 

\textbf{Correctness.}
In the solution of MILP \ref{milp-pbm1-simple}, $x_{ir}=1$ iff $\tau_i$ is executed in round $r$; and $y_r$ indicates whether the round $r$ is used by any transaction. The objective is to minimize the number of rounds used, minimizing total execution time (makespan).

Constraint~\eqref{con:p-cores} enforces the per-round parallelism limit. 
Since the binary variable $x_{ir}$ indicates that transaction~$\tau_i$ executes in round $r$,
the sum $\sum_i x_{ir}$ counts the number of transactions assigned to that round. 
At most $p$ cores are available, and hence we require $\sum_i x_{ir} \le p \cdot y_r$, where $y_r = 1$ only if round~$r$ is actually used. 
Therefore, no round can host more than $p$ concurrent transactions, and unused rounds ($y_r = 0$) cannot schedule any work.
Observe that in this case, the solution does not need to assign transactions to specific cores.

Constraint~\eqref{con:order} enforces precedence. 
For every directed edge $(i,j) \in E$, transaction~$\tau_i$ must complete before~$\tau_j$ begins. 
The term $\sum_{t=1}^r x_{jt}$ equals~1 if $\tau_j$ is scheduled in a round $\le r$, while $\sum_{t=1}^{r-1} x_{it}$ equals~1 if $\tau_i$ is scheduled strictly before round~$r$. 
The inequality
\[
\sum_{t=1}^r x_{jt} - \sum_{t=1}^{r-1} x_{it} \le 0, \quad \forall r,
\]
therefore excludes any assignment in which $\tau_j$ is placed before $\tau_i$. 
Together, constraints~\eqref{con:p-cores} and~\eqref{con:order} ensure that (i) no more than $p$ transactions execute in parallel in any round and (ii) all precedence edges are respected in the resulting schedule.

\subsubsection{Heterogeneous Transactions}

We now extend the model to the heterogeneous case, in which a block contains both simple and complex transactions with different execution times. MILP~\ref{milp-pbm1-simple&complex} encodes this mixed setting. The total execution time (makespan) found is denoted by $M$, and $B$ is a parameter that must be set to be larger than any possible $M$. The objective is to minimize the makespan $M$ subject to the assignment, timing, and ordering constraints.

Each transaction $\tau_i$ is assigned to exactly one core via a binary variable $x_{ic}$ and scheduled with start and end times $s_i$ and $e_i$, respectively.  Precedence edges $(i,j) \in E$ enforce dependency order via \Cref{con:order-c1}, ensuring that conflicting transactions follow the sequence order. For non-conflicting pairs, overlap is permitted unless the two transactions are assigned to the same core. Same-core assignment is captured by the auxiliary variable $w_{ijc}$ (the logical AND of $x_{ic}$ and $x_{jc}$), and $z_{ij}$ indicates whether the pair shares any core. If $z_{ij}=1$, the binary selector $y_{ij}$ determines their order: $y_{ij}=1$ means $\tau_i$ precedes $\tau_j$; otherwise, $y_{ij}=0$ means the reverse. 
In \Cref{con:order1} and \Cref{con:order2}, observe the following cases:
\begin{itemize}
    \item $z_{ij}=0$: the constraints have no effect due to the runtime limit term $B$.
    \item $z_{ij}=1, y_{ij}=1$:  \Cref{con:order2} has no effect, only \Cref{con:order1} applies, yielding $e_i \le s_j$.
    \item $z_{ij}=1, y_{ij}=0$: \Cref{con:order1} has no effect, only \Cref{con:order2} applies, yielding $e_j \le s_i$.
\end{itemize}
These conditional constraints ensure that same-core transactions are serialized.

\textbf{Correctness.}
Constraint~\eqref{con:onetrans-c1} ensures that every transaction is assigned to exactly one core. 
Precedence edges $(i,j) \in E$ are enforced by~\eqref{con:order-c1}, which requires that any dependent transaction $\tau_j$ starts only after $\tau_i$ has completed. 
Together, these constraints preserve the total order implied by the dependency graph.
For non-conflicting pairs, the coupling of $w_{ijc}$, $z_{ij}$, and $y_{ij}$ ensures that same-core pairs are serialized by 
\eqref{con:order1}--\eqref{con:order2}, while pairs on different cores remain unconstrained. 
Together, these properties ensure that MILP \ref{milp-pbm1-simple&complex} produces a feasible schedule that respects all dependencies and minimizes the overall makespan~$M$.

\subsubsection{Complexity} 

Both MILP formulations grow quickly with the number of transactions $n$ and cores $p$. 
The homogeneous case (MILP \ref{milp-pbm1-simple}) involves $\mathcal{O}(nR)$ binary variables and $\mathcal{O}(nR + |E|R)$ constraints, while the heterogeneous case (MILP~\ref{milp-pbm1-simple&complex}) expands to $\mathcal{O}(n^2p)$ binary variables, $\mathcal{O}(n)$ integer variables, and $\mathcal{O}(n^2p + |E|)$ constraints. 
Each formulation generalizes well-known NP-hard multiprocessor scheduling problems, implying that exact solutions become computationally prohibitive as $n$ and $p$ increase. 

    \begin{algorithm}[H]
      \caption{$\builddag{}$. Recall the conflict definition from \Cref{subsubsec:conflicts}}
      \label{heu:build-DAG}%
      \footnotesize
      \begin{algorithmic}[1]
        \Function{$\builddag{T}$}{}
          \State $V \gets \{\tau:\tau \in T\}$; $E \gets \emptyset$
          \For{\textbf{each} $i \in [1..n-1]$}
             \For{\textbf{each} $j \in [i+1..n]$}
                \If{$\mathrm{conflict}(\tau_i,\tau_j)$} $E \gets E \cup \{(\tau_i,\tau_j)\}$
                \EndIf
            \EndFor
          \EndFor
          \State \textbf{return} $G=(V,E)$
      \EndFunction
      \end{algorithmic}
      \end{algorithm}
          
    
    
            


    \begin{algorithm}[H]
      \caption{$\preprocess{}$}%
      \label{heu:preprocessing_problem1}%
      \footnotesize
      \begin{multicols}{2}
      \begin{algorithmic}[1]
      \Function{$\preprocess{G}$}{}
        \For{\textbf{each} $\tau \in V$}
            \State $\successor{[\tau]} \gets \{\tau' : (\tau, \tau') \in E\}$
            \State $\predecessor{[\tau]} \gets \{\tau' : (\tau', \tau) \in E\}$
            \State $\outdegree{[\tau]} \gets | \successor{[\tau]} |$
            \State $\indegree{[\tau]} \gets | \predecessor{[\tau]} |$
        \EndFor
        \State $Q \gets \{\tau: \outdegree{[\tau]}=0\}$ \Comment{Set of sinks}
        \While{$Q \neq \emptyset$}
            \State $\tau \gets$ any element in $Q$
            \State $Q \gets Q \setminus \{\tau\}$
            \State $\height{[\tau]} \gets \executiontime{\tau}+\max_{\tau' \in \successor{[\tau]}}\{\height{[\tau']}\}$
            \State $\volume{[\tau]} \gets \executiontime{\tau}+\sum_{\tau' \in \successor{[\tau]}}\{\volume{[\tau']}\}$
            \For{\textbf{each} $\tau' \in \predecessor{[\tau]}$}
                \State $\outdegree{[\tau']} \gets \outdegree{[\tau']} - 1$
                \If{$\outdegree{[\tau']} = 0$} $Q \gets Q \cup \{\tau'\}$
                \EndIf
            \EndFor
        \EndWhile
        \For{\textbf{each} $\tau \in V$}
            \State $\priority{[\tau]} \gets (-\height{[\tau]},-\volume{[\tau]},-|\successor{[\tau]}|,\id{\tau})$
        \EndFor
        \State \textbf{return}$(\priority,\indegree,\successor)$
      \EndFunction
      \end{algorithmic}
      \end{multicols}
      \end{algorithm}


\begin{algorithm}[H]
\caption{$\findschedulePone{}$}%
\label{heu:scheduling_problem1}%
\footnotesize
\begin{multicols}{2}
\begin{algorithmic}[1]
\Function{$\findschedulePone{T, p, \priority, \indegree, \successor}$}{}
    \State $now \gets 0$
    \State $free \gets [1..p]$ \Comment{Free cores}
    \State $running \gets \emptyset$ \Comment{Trans.~that are running}
    \State $ready \gets \{\tau \in T: \indegree{[\tau]}=0\}$ \Comment{Ready tx}
    \Repeat
        \While{$(ready \neq \emptyset) \land (free \neq \emptyset)$}
            \State $\tau \gets \arg \min \{ \priority{[\tau']}: \tau' \in ready \}$
            \State $ready \gets ready \setminus \{\tau\}$
            \State $running \gets running \cup \{\tau\}$
            \State $\start{[\tau]} \gets now$
            \State $\finish{[\tau]} \gets now + \executiontime{\tau}$
            \State $\core{[\tau]} \gets$ any element in $free$
            \State $free \gets free \setminus \{\core{[\tau]}\}$
        \EndWhile
        \State $now \gets \min\{\finish{[\tau]} : \tau \in running\}$
        \State $S \gets \{\tau \in running \;:\; \finish{[\tau]} = now\}$
        \State $running \gets running \setminus S$
        \For{\textbf{each} $\tau \in S$}
        \For{\textbf{each} $\tau' \in \successor{[\tau]}$}
            \State $\indegree{[\tau']} \gets \indegree{[\tau']} - 1$
            \If{$\indegree{[\tau']} = 0$} 
                \State $ready \gets ready \cup \{\tau'\}$
            \EndIf
        \EndFor
        \State $free \gets free \cup \{\core{[\tau]}\}$
        \EndFor
    \Until{$(running = \emptyset) \land (ready = \emptyset)$}
    \State \textbf{return} $(\start, \core)$
\EndFunction
\end{algorithmic}
\end{multicols}
\end{algorithm}

  \begin{algorithm}[H]
  \footnotesize
    \caption{OBS Heuristics}
    \label{heu:problem1}
    \begin{algorithmic}[1]
        \State \textbf{Input:} 
        Block $T=(\tau_1,\dots,\tau_n)$.  
        Number of cores  $p$.
        \State \textbf{Output:} A schedule with the start time $\start{[\tau]}$ and core $\core{[\tau]}$ assigned to each transaction $\tau \in T$.


      \Statex
      \State $G \gets \builddag{T}$
      \State $(\priority,\indegree, \successor) \gets \preprocess{G}$
      \State $(\start, \core) \gets \findschedulePone{T, p, \priority, \indegree, \successor}$
    \end{algorithmic}
  \end{algorithm}

\subsection{Heuristics}

Our heuristics (\Cref{heu:build-DAG,heu:preprocessing_problem1,heu:scheduling_problem1,heu:problem1}, 
use a conflict-aware greedy scheduler, guided by transaction priorities. It proceeds in three stages:

    \noindent
    \textbf{(1) DAG construction.} For the given block $T=(\tau_1,\dots,\tau_n)$, we build a dependency DAG $G=(V,E)$ (\Cref{heu:build-DAG}). Each node is a transaction $\tau \in T$, and we add an edge $(\tau_i,\tau_j)$ whenever $\tau_i$ and $\tau_j$ conflict (as defined in \Cref{subsubsec:conflicts}) and $i<j$. This DAG encodes all precedence constraints that must be respected at execution time.
    
       \noindent
       \textbf{(2) Priority scoring.} We then extract structural information from $G$ (\Cref{heu:preprocessing_problem1}). For each transaction $\tau$, we compute:
    \begin{itemize}
        \item $\height{[\tau]}$: the critical-path length rooted at $\tau$ (i.e., $\tau$'s own execution time plus the maximum downstream height),
        \item $\volume{[\tau]}$: the total execution volume of $\tau$ and its descendants,
        \item $|\successor{[\tau]}|$: the fan-out of $\tau$.
    \end{itemize}
    A single backward pass over the DAG computes these values. We assign each transaction a deterministic priority key
    $
        \priority{[\tau]} := \big(-\height{[\tau]},\; -\volume{[\tau]},\; -|\successor{[\tau]}|,\; \id{\tau}\big),
    $
    which favors long critical paths first, then heavier subtrees, then higher fan-out, and $\id{\tau}$ breaks ties.
    
    \noindent
    \textbf{(3) Transaction scheduling.} Finally, we simulate the block execution on $p$ cores (\Cref{heu:scheduling_problem1}). The scheduler maintains (i) a set of \emph{ready} transactions (initially those with in-degree zero in $G$), (ii) a set of \emph{running} transactions with assigned cores and finish times, and (iii) the current time $now$. Whenever any core becomes free, we schedule the highest-priority ready transaction to that core. When a transaction finishes, we decrement the in-degree of its successors and add any successor with zero in-degree to the set of ready transactions. This event-driven loop continues until all transactions are scheduled. \Cref{heu:problem1} combines these steps into a complete pipeline.

\textbf{Complexity.}
The DAG construction in \Cref{heu:build-DAG} compares each ordered pair $(\tau_i,\tau_j)$ with $i<j$, and is therefore $\mathcal{O}(n^2)$ in the worst case.\footnote{In practice, this can often be pruned with sparse access lists, but we analyze the dense upper bound.} The preprocessing pass in \Cref{heu:preprocessing_problem1} performs a single reverse topological traversal plus constant work per edge, i.e., $\mathcal{O}(n + |E|)$. The scheduler in \Cref{heu:scheduling_problem1} runs as an event-driven simulation with at most $n$ schedule events and $n$ completion events; using a priority queue for the ready set, this is $\mathcal{O}\big(n \log n + |E|\big)$. Overall, the heuristic end-to-end runtime is dominated by DAG construction and is $\mathcal{O}(n^2)$ in the worst case.
Regarding the approximation ratio, since the heuristic implements a list scheduling, we know that it gives a
schedule with makespan at most $2-1/p$ times the optimal.

\section{Parallel-Block Construction (PBC)} \label{sec:problem2}
We now turn to the block-construction problem. Given a
mempool $\mempool=\{\tau_1, \ldots, \tau_n\}$ of pending transactions, 
we construct an undirected conflict graph $G=(\mempool,E)$ whose nodes correspond to the transactions in $\mempool$, with an edge $(\tau_i,\tau_j)\in E$ if and only if the two transactions conflict (as defined in \Cref{subsubsec:conflicts}).
The block runtime limit~$B$ and the conflicts must be respected in the schedule obtained for the transactions selected from $T$ to be included in the block.
This setting combines both \emph{selection} (which transactions to include) and \emph{scheduling} (how to assign them across cores) under dependency and time budget constraints.

\subsection{PCB MILP Formulations}

\subsubsection{Homogeneous Transactions}

\begin{figure}[t!]
\centering
\begin{tcolorbox}[enhanced, arc=1mm, boxrule =0.8pt, colback=white, colframe=black, width=\textwidth, left =1pt, right=6pt,  ams align,fontupper=\footnotesize,]
\max \quad & \sum_{i \in [n]} \sum_{r \in [R]} w_i\, x_{ir} 
    && && \text{(Maximize total weight)} \label{obj:max-s2} \\[0.5ex]
\text{s.t.}\quad
& \sum_{r \in [R]} x_{ir} \leq 1 
    && \forall i \in [n]  
    && \text{(Each trans.~scheduled at most once)} \label{con:onetrans-s2} \\
& \sum_{i \in [n]} x_{ir} \leq p 
    && \forall r \in [R]
    && \text{(At most $p$  transactions per round)} \label{con:p-cores-s2} \\
& x_{ir} + x_{jr} \leq 1 
    && \forall (i,j) \in E, i<j,\; \forall r \in [R]  
    && \text{(No conflicting pair in same round)} \label{con:conflict-s2} \\
& x_{ir} \in \{0,1\} 
    && \forall i \in [n],\; \forall r \in [R] 
    && \text{(Binary assignment)} \label{con:binary}
\end{tcolorbox}
\captionsetup{labelformat=simple,labelsep=colon,name={MILP},justification=centering}
\caption{Parallel Block Construction - Homogeneous Transactions}
\label{milp-pbm2-simple}
\end{figure}

We first consider the homogeneous case, where all transactions have identical execution time~$t$. The block gas budget $B$ therefore allows at most  $R=\lfloor B/t \rfloor$ execution rounds. MILP~\ref{milp-pbm2-simple} formalizes the selection-and-scheduling problem for this setting. Here, $x_{ir}=1$ if and only if transaction~$\tau_i$ is selected and executed in round~$r$, yielding a validator reward $w_i := \reward{\tau_i}$ for its inclusion in the block.  

The problem can be viewed as a weighted packing task: we fill each of the $R$ rounds with up to $p$ non-conflicting transactions, choosing the subset that maximizes total weight (i.e., reward $\sum_{i} w_i$)  while respecting both conflict and capacity constraints. 

\textbf{Correctness.}
Constraint~\eqref{con:onetrans-s2} ensures that each transaction is scheduled at most once across all rounds.  
Constraint~\eqref{con:p-cores-s2} enforces the validator’s core capacity, limiting each round to at most $p$ concurrent transactions.  
Constraint~\eqref{con:conflict-s2} excludes conflicting pairs from being assigned to the same round, preserving deterministic execution semantics.  
Together, these constraints guarantee that every feasible solution corresponds to a conflict-free parallel schedule that fits within the gas budget.  
The objective~\eqref{obj:max-s2} then selects the optimal subset of transactions that maximizes total validator reward under these feasibility constraints.

\subsubsection{Heterogeneous Transactions}

\begin{figure*}[t!]
\begin{tcolorbox}[enhanced, arc=1mm, boxrule =0.8pt, colback=white, colframe=black, width=\textwidth, left =1pt, right=6pt,  ams align,fontupper=\footnotesize,]
\max \ & \sum_{i \in [n]}  w_i\, v_i 
    && \text{(Maximize total weight)} \label{obj:max-c3} \\[0.5ex]
\text{s.t.}\ 
& \sum_{c \in [p]} x_{ic} = v_i
    && \forall i \in [n]  
    && \text{(One core max)} \label{con:one-assignment-c3} \\
& e_i - s_i = t_i v_i 
    && \forall i: i \in [n] ,
    && \text{(Time activation )} \label{con:noneg-c3} \\
& 0 \leq s_i \leq B v_i, 0 \leq e_i \leq B v_i
    && \forall i: i \in [n] , 
    && \text{(Activation bounds)}  \\
& e_i - s_j \leq B \bigl( 1 - y_{ij} \bigr),
    && \forall (i,j): (i,j) \in E \land i<j, 
    && \text{(Conflicts)} \label{con:conflict1-c3} \\
& e_j - s_i \leq B \bigl( y_{ij}  \bigr),
    && \forall (i,j): (i,j) \in E \land i<j, 
    && \text{(Conflicts)} \label{con:conflict2-c3} \\ 
&w_{ijc} \leq x_{ic}, w_{ijc} \leq x_{jc} 
    &&  \forall (i,j) \in \mathcal{P}_{\text{non}} , \forall c \in [p]\\
&w_{ijc} \geq x_{ic} + x_{jc} -1 
    &&  \forall (i,j) \in \mathcal{P}_{\text{non}}, \forall c \in [p]\\
&z_{ij} =\sum_{c \in [p]} w_{ijc} 
    &&  \forall (i,j) \in \mathcal{P}_{\text{non}}
    && \text{(Same core flag)}\\
&e_i \le s_j + B \bigl( (1 - y_{ij}) + (1-z_{ij}) \bigr), 
    && \forall (i,j) \in \mathcal{P}_{\text{non}}\label{con:order1-c2}\\
&e_j \le s_i + B \bigl( y_{ij} + (1-z_{ij}) \bigr), 
    && \forall (i,j) \in \mathcal{P}_{\text{non}}\label{con:order2-c2}\\
& s_i, e_i \in \mathbb{N} 
    && \forall i \in [n] 
    && \text{(Integer assignment)} \label{con:linear-assign-c3} \\
& v_i,w_{ijc} x_{ic}, y_{ij}, z_{ij} \in \{0,1\}
    && \forall i,j \in [n],\; \forall c \in [p] 
    && \text{(Binary assignment)} \label{con:binary-assign-c3}
\end{tcolorbox}
\captionsetup{labelformat=simple,labelsep=colon,name={MILP}, justification=centering}
\caption{Parallel Block Construction - Heterogeneous Transactions. Let $\mathcal{P}_{\text{non}}:=\{ (i,j) : i<j \land (i,j) \notin E \land (j,i) \notin E \}$ be the set of pairs of transactions without conflicts.}
\label{milp-pbm2-simple&complex}
\end{figure*}

We now extend to the case of transactions with heterogeneous execution times~$t_i$. 
MILP~\ref{milp-pbm2-simple&complex} jointly models transaction selection and parallel scheduling under conflicts, core capacity, and runtime constraints.  The binary variable $v_i$ determines whether transaction $\tau_i$ is selected to be included in a block. If selected ($v_i=1$), it must be assigned to exactly one core via~$x_{ic}$ and executed during the time window $[s_i, e_i]$ with duration $t_i$. Conflict edges $(i,j)\in E$ forbid temporal overlap between the conflicting transactions. Their mutual ordering is decided by $y_{ij}$, enforced through constraints \eqref{con:conflict1-c3}–\eqref{con:conflict2-c3}. For non-conflicting pairs, overlap is allowed unless they share a core. Same-core assignment is captured by $w_{ijc}$, the logical AND of $x_{ic}$ and $x_{jc}$, and $z_{ij}$ which flags whether a pair shares any core.
If $z_{ij}=1$, the pair must be serialized, with order selected by $y_{ij}$ as enforced in constraints \eqref{con:order1-c2}–\eqref{con:order2-c2}. The objective~\eqref{obj:max-c3}  maximizes the total weight $\sum_{i \in [n]} w_i v_i$ subject to gas and core limits, balancing which transactions are included and how they are scheduled.

\textbf{Correctness.} 
Constraint~\eqref{con:one-assignment-c3} ensures that every selected transaction occupies exactly one core, while \eqref{con:noneg-c3} and the activation bounds couple start and end times to the selection variable~$v_i$.  
Conflict pairs $(i,j)\in E$ are serialized via constraints~\eqref{con:conflict1-c3}–\eqref{con:conflict2-c3}, preventing overlapping execution regardless of order.  
For non-conflicting transactions, same-core co-assignment is detected by $w_{ijc}$ and summarized by $z_{ij}$; if $z_{ij}=1$, constraints~\eqref{con:order1-c2}–\eqref{con:order2-c2} enforce a valid ordering determined by $y_{ij}$.  
These constraints guarantee that all selected transactions can be executed within the gas limit~$B$ without violating conflict or resource dependencies.  
The objective then identifies the highest-reward feasible subset and schedule.

\subsubsection{Complexity} Both the homogeneous (MILP~\ref{milp-pbm2-simple}) and heterogeneous (MILP~\ref{milp-pbm2-simple&complex}) formulations of PCB  generalize NP-hard problems such as weighted parallel-machine scheduling and multidimensional knapsack.  
The homogeneous formulation involves $\mathcal{O}(nR)$ binary variables and $\mathcal{O}(nR + |E|R)$ constraints, while the heterogeneous one expands to $\mathcal{O}(n^2p)$ binary variables and $\mathcal{O}(n^2p + |E|)$ constraints.  
Solver time thus grows sharply with block size and heterogeneity, motivating the use of fast approximation methods. 

\subsection{Heuristics}

\begin{algorithm}[t!]
\caption{$\findschedulePtwo{}$ }
\label{heu:p2-schedule}
\footnotesize
\begin{multicols}{2}
\begin{algorithmic}[1]
\Function{$\findschedulePtwo{T, p, B, \priority, \conflicts}$}{}
  \State $now \gets 0$
  \State $free \gets [1..p]$ \Comment{Free cores}
  \State $running \gets \emptyset$ \Comment{Trans. running}
  \State $\selected \gets \emptyset$ \Comment{Trans. completed}
  \State $ready \gets T$ \Comment{Pending transactions}
    \State $\deferred \gets \emptyset$ \Comment{Conflicting with running}
  \Repeat
    \While{$(ready \setminus \deferred \neq \emptyset) \land (free \neq \emptyset)$}
        \State $\tau \gets \arg \min \{ \priority{[\tau']} :$ \\
        \hspace{3cm} $\tau' \in ready \setminus \deferred \}$
        \If{$(running \cap \conflicts{[\tau]} \neq \emptyset)$}
            \State $\deferred \gets \deferred \cup \{\tau\}$
            \State \textbf{continue}
        \EndIf
        \State $ready \gets ready \setminus \{\tau\}$
        \If{$now + \executiontime{\tau} > B$}
        \State \textbf{continue}
      \EndIf
      \State $running \gets running \cup \{\tau\}$
      \State $\selected \gets \selected \cup \{\tau\}$
      \State $\start{[\tau]} \gets now$
      \State $\finish{[\tau]} \gets now + \executiontime{\tau}$
      \State $\core{[\tau]} \gets$ any element in $free$
      \State $free \gets free \setminus \{\core{[\tau]}\}$
    \EndWhile
    \State $now \gets \min\{\finish{[\tau]} : \tau \in running\}$
    \State $S \gets \{\tau \in running \;:\; \finish{[\tau]} = now\}$
    \State $running \gets running \setminus S$
    \For{\textbf{each} $\tau \in S$}
        \State $free \gets free \cup \{\core{[\tau]}\}$
        \State $\deferred \gets \deferred \setminus \conflicts{[\tau]}$
    \EndFor

    \Until{$(running = \emptyset) \land (ready = \emptyset)$}
    \State \textbf{return } $(\selected,\start,\core)$
\EndFunction
\end{algorithmic}
\end{multicols}
\end{algorithm}

Our mempool heuristics (\Cref{heu:p2-score,heu:p2-schedule,heu:problem2}) greedily assemble a conflict-free parallel schedule of transactions under the runtime limit $B$. 
They combine lightweight scoring and event-driven dispatch to approximate the MILP objective efficiently.


    \noindent
    \textbf{Priority Scores:} For each transaction $\tau$ in the mempool, we compute its value density as $\rho[\tau] \coloneqq \frac{\reward{\tau}}{\executiontime{\tau}}$
     and $\degree{[\tau]}$ is the number of transactions with which $\tau$ conflicts.
    The composite priority key 
    $\priority{[\tau]} \coloneqq ( -\rho[\tau],\degree{[\tau]},-\reward{\tau},\id{\tau}),$
    favors high-reward, low-conflict transactions, breaking ties deterministically by~$\id{\tau}$.

    \noindent
    \textbf{Transaction Scheduling:} Using the priorities above, we iteratively dispatch transactions to free cores (\Cref{heu:p2-schedule}). At each step, the scheduler maintains three sets: \emph{ready} transactions (eligible for execution), \emph{running} transactions (currently executing), and a \emph{deferred} set of candidates that conflict with running ones. When a core becomes free, the highest-priority non-conflicting transaction is launched, provided the cumulative runtime used remains within~$B$. On each completion event, the corresponding transaction is removed from \emph{running}, its conflicts are cleared from \emph{deferred}, and its core is released.  
    This process continues until either the runtime budget or transaction pool is exhausted.  
    \Cref{heu:problem2} integrates both steps into a full end-to-end pipeline.

\textbf{Complexity.}  The conflict graph construction has complexity $\mathcal{O}(n^2)$.
Computing transaction degrees and priorities has also complexity $\mathcal{O}(n^2)$.  
The event-driven scheduler then performs $\mathcal{O}(n \log n)$ priority operations plus $\mathcal{O}(np)$ updates for dispatch and completion across $p$ cores.  
Overall, the heuristic runs in polynomial time, dominated by graph construction and conflict detection.

\begin{figure*}[t!]
\centering

\begin{minipage}[t]{0.49\textwidth}

\input{Heuristics/pbm2_scoring}

\end{minipage}\hfill
\begin{minipage}[t]{0.49\textwidth}

\input{Heuristics/pbm2_main}

\end{minipage}

\end{figure*}

\section{Experiments} \label{sec:experiments}

As described, we study two validator-side tasks: (i) OBS: scheduling a fixed ordered block on $p$ cores to minimize makespan and (ii) PBC: selecting and scheduling mempool transactions to maximize reward under a runtime budget $B$. We evaluate our MILP formulations and conflict-aware greedy heuristics on transaction traces extracted from the Ethereum mainnet for blocks 21,631,019 - 21,635,079, spanning 15 - 16 January 2025. 
Each transaction trace records the execution outcome, including the read and write sets, gas used, and internal call tree produced by the Ethereum Virtual Machine (EVM).  
From these traces, we extract per-transaction \texttt{gasUsed}, \texttt{to}/\texttt{from} addresses, storage keys accessed, and all contract calls.  
Conflicts are then derived exactly from overlapping read/write sets (as described in \Cref{subsubsec:conflicts}). In addition to comparing our heuristics against the MILP baselines, for OBS, we include a baseline inspired by Solana’s declared-access execution model (Sealevel), where transactions expose read/write account sets up front and the runtime schedules only non-conflicting transactions to run concurrently~\cite{solanaSealevel,yakovenko2017solana}. We denote this baseline by \textbf{Sol}. On our Ethereum-derived instances (using the access sets extracted from traces), we implement Sol as a deterministic, event-driven scheduler that scans transactions in block order and dispatches each transaction as soon as a core becomes available and it does not conflict with currently running transactions. For PBC, we instead compare against a simple \textbf{reward-greedy} baseline (\textbf{RG}) that uses the same non-conflicting dispatch rule but selects candidates purely by reward-based priority (i.e., favoring higher-fee transactions, breaking ties by arrival order), yielding a lightweight selection-and-scheduling baseline under the conflict constraints.

\subsection{Hardware and Software}

All experiments were run on a Dell PowerEdge R7615 equipped with a single AMD EPYC 9654P (96 cores/ 192 threads) and $\approx$1.5TiB RAM, running Ubuntu 24.04.3 LTS. We use MATLAB R2025b for the heuristics and invoke HiGHS 1.7.1 through MATLAB’s \texttt{intlinprog} for MILPs. We do not limit the number of threads: MATLAB and HiGHS may use all available cores. Unless otherwise specified, solver tolerances follow MATLAB defaults. We set a maximum running time of $\mathit{MaxTime}:=6000$ s.~for the solver. All heuristics are deterministic; repeated runs on the same workload yield identical schedules.

\subsection{Workload Construction}
Our experiments use real Ethereum transactions to synthesize controlled workloads for both problems.  
To control workload size across configurations $(R \text{ or } B, p, X)$, we build synthetic workloads by taking consecutive mainnet transactions in their original order and re-aggregating them, rather than relying on variable on-chain block boundaries.

\noindent\textbf{Trace filtering.}  We parse transactions and separate them into two categories:
\begin{itemize}
  \item \textbf{Homogeneous} transactions: pure ETH transfers that do not invoke any smart contract, and therefore use the default gas amount of $21,000$ units.
  \item \textbf{Heterogeneous} transactions: the full set of observed transactions, including contract calls with varying performed computations.
\end{itemize}
Each transaction’s gas used $\gasused{\tau}$ is taken directly from its execution trace, and its execution time $\executiontime{\tau}$ is normalized proportionally to gas use.

\noindent\textbf{OBS workloads.}  
For each experiment, we collect a sequence of transactions (Homogeneous or Heterogeneous) in their original blockchain order, ignoring actual block boundaries.  
We then aggregate this sequence into blocks of size determined by either
  (a) the number of rounds $R$ for homogeneous cases, each round able to process up to $p$ transactions; or
  (b) the gas budget $B$ for heterogeneous cases, stopping once $\sum_{\tau} \gasused{\tau} \approx B$.
For each configuration $(R\text{ or }B,p)$, we construct five consecutive, non-overlapping workloads.
This corresponds to simple transactions extracted from 219 blocks ($9,600$ transactions) of mainnet data for the homogeneous case, and 14 blocks ($2,460$ transactions) of mainnet data for the heterogeneous.

\noindent\textbf{PBC workloads.}  
To emulate a validator’s mempool, we use a sliding window of transactions.  
Given configuration $(R,p,X)$ or $(B,p,X)$, where $X$ is the \emph{pool factor}, we select a candidate pool containing approximately $X$ times the transactions (or total gas) that fit in one block:
\begin{itemize}
  \item 
  Homogeneous case: $R \cdot p \cdot X$ transactions;
  \item 
  Heterogeneous case: $B \cdot p \cdot X$, where $\sum_{\tau}\gasused{\tau} \approx B$.
\end{itemize}
The corresponding MILP and heuristic then select and schedule a feasible subset for inclusion.  
For each configuration, we again construct five independent workloads.  
In practice, this corresponds to simple transactions extracted from $2,989$ blocks ($137,200$ transactions) of mainnet data for the homogeneous case, and 24 blocks ($4,280$ transactions) of mainnet data for the heterogeneous case.

\noindent\textbf{Experiment parameters.}  
For homogeneous transactions, experiment duration is discretized into equal-length rounds, so progress is naturally parameterized by the number of rounds \(R\), with per-round capacity \(p\). For heterogeneous transactions, we parameterize the experiment duration by a continuous runtime limit \(B\), which directly captures feasibility under varying processing times. For homogeneous cases, we evaluate each configuration with rounds \(R\in\{10,30,100\}\), cores \(p\in\{2,4,8\}\), and (when applicable) pool factor \(X\). For mempool experiments we use \(X\in\{2,4,8\}\) in the homogeneous case and \(X\in\{2,4\}\) in the heterogeneous case. For heterogeneous cases, we set the runtime limit to \(B\in\{210\text{K},630\text{K},2100\text{K}\}\). For every R \text{ or }B,p,X configuration, we run five 
non-overlapping workloads and report the mean. 


\subsection{Hypotheses} \label{subsec:hypotheses}

We evaluate the following hypotheses:

  \noindent
  \textbf{H1 (MILP scaling):} MILP runtime grows steeply with problem size and heterogeneity.

  \noindent
  \textbf{H2 (Heuristic speed):} The conflict-aware greedy heuristics runs orders of magnitude faster than MILPs across all settings.

  \noindent
  \textbf{H3 (Heuristic quality):} 
  The heuristics yield parallel schedules that approach the performance of MILP solutions.

  \noindent
  \textbf{H4 (Practical speedup):} The heuristics yield parallel schedules with significant speedup with respect to sequential executions.


\section{Results and Discussion} \label{sec:results}

We now present results for OBS and PBC under homogeneous and heterogeneous workloads. Unless stated otherwise, each reported value is the mean over five independent instances per configuration. We compare our conflict-aware greedy heuristic (GH) to the MILP baselines and to the additional baselines introduced in \Cref{sec:experiments}.

\subsection{Ordered-Block Scheduling}

We first compare the runtime of the MILP solver against GH and Sol for OBS on homogeneous workloads with $p \in \{2,4,8\}$ and $R \in \{10,30,100\}$. 
Solver time increases rapidly with both $R$ and $p$, growing from milliseconds on small instances to thousands of seconds for $R{=}100$, $p{=}8$, confirming the expected combinatorial scaling (\textbf{H1}). GH and Sol, by contrast, remain consistently below $0.1$\,s even at the largest configuration (\textbf{H2}). \Cref{tab:pbm1_simple} shows that on homogeneous instances, GH matches the MILP solution across all tested $(R,p)$ (\textbf{H3}). Sol is close but degrades on larger instances; for example, at $R{=}100$, Sol requires $106.2$ rounds for $p{=}2$ and $118.4$ rounds for $p{=}4$, compared with $100$ rounds for MILP/GH. This indicates that simple in-order dispatch leaves avoidable stalls even when conflicts are known. GH achieves a speedup similar to that of MILP, which is near-linear for these low-conflict homogeneous transfer workloads. This gap reflects the difference between a reactive in-order dispatcher (Sol), which may block on transient conflicts, and our proactive scheduling approach (GH), which pre-computes a conflict-respecting schedule that smooths execution by prioritizing transactions on the critical path.

We observe a similar runtime separation on heterogeneous workloads, parameterized by gas budget $B \in \{210K,630K,2100K\}$. 
As $B$ grows, solver runtime escalates by orders of magnitude with both $p$ and $B$ (\textbf{H1}), exceeding our $6{,}000$\,s execution time limit for the largest configuration.  GH and Sol, by contrast, remain sub-second throughout (\textbf{H2}). \Cref{tab:pbm1_complex} confirms that GH's makespans differ from the MILP’s best integer solutions by at most a few percent (and in some cases even improve upon the solver’s incumbent when the solver exceeds our runtime limit) and also consistently outperforms Sol (\textbf{H3}). Overall speedups over 1-core sequential execution are in the $\sim$1.6--2.3$\times$ range (\Cref{tab:pbm1_complex}), confirming \textbf{H4} but also highlighting a key limitation of OBS: even with many cores, a fixed consensus order and conflict-induced dependencies cap achievable parallelism.

\begin{table}[t!]
\centering
\tiny
\begin{tabular}{|c|c|c|c|c|c|c|}
\hline
      & \multicolumn{2}{c|}{\textbf{R = 10}} 
      & \multicolumn{2}{c|}{\textbf{R = 30}} 
      & \multicolumn{2}{c|}{\textbf{R = 100}} \\ \hline
\textbf{p} 
      & \textbf{MILP/GH} & \textbf{Sol} 
      & \textbf{MILP/GH} & \textbf{Sol} 
      & \textbf{MILP/GH} & \textbf{Sol} \\ \hline
\textbf{2} 
      & 10/2      & 10.2/1.96 
      & 30/2      & 30.2/1.99 
      & 100/2     & 106.2/1.88 \\ \hline
\textbf{4} 
      & 11/3.64   & 13.2/3.03 
      & 30.6/3.92 & 38/3.16   
      & 100/4     & 118.4/3.38 \\ \hline
\textbf{8} 
      & 11.6/6.9  & 15.2/5.26 
      & 31.6/7.6  & 44.6/5.38 
      & 100/8     & 115.2/6.94 \\ \hline
\end{tabular}
\caption{OBS - Homogeneous transactions: Average number of rounds and average speedup (related to the sequential runtime) for MILP solver and greedy heuristic (identical results) compared to Solana (Sol).}
\label{tab:pbm1_simple}
\end{table}

\begin{table}[t!]
\centering
\tiny
\begin{tabular}{|c|cc|cc|cc|}
\hline
           & \multicolumn{2}{c|}{\textbf{B = 210K}}          & \multicolumn{2}{c|}{\textbf{B = 630K}}          & \multicolumn{2}{c|}{\textbf{B = 2100K}}         \\ \hline
\textbf{p} & \multicolumn{1}{c|}{\textbf{GH}} & \textbf{Sol} & \multicolumn{1}{c|}{\textbf{GH}} & \textbf{Sol} & \multicolumn{1}{c|}{\textbf{GH}} & \textbf{Sol} \\ \hline
\textbf{2} & \multicolumn{1}{c|}{100.49/1.65} & 103.69/1.59  & \multicolumn{1}{c|}{100.29/1.57} & 107.56/1.45  & \multicolumn{1}{c|}{$98.68^{\dagger}$/1.81}  & $117.15^{\dagger}$/1.51  \\ \hline
\textbf{4} & \multicolumn{1}{c|}{100/1.82}    & 102.9/1.77   & \multicolumn{1}{c|}{100/1.98}    & 105/1.86     & \multicolumn{1}{c|}{100/2.15}    & 111.03/1.94  \\ \hline
\textbf{8} & \multicolumn{1}{c|}{100/2.26}    & 101.19/2.22  & \multicolumn{1}{c|}{100/2.03}    & 101.13/2     & \multicolumn{1}{c|}{$77.67^{\dagger}$/2.29}  & $81.77^{\dagger}$/2.17   \\ \hline
\end{tabular}
\caption{OBS - Heterogeneous transactions: Heuristics solution makespan $M$ as a percentage of the optimal solution found by the MILP solver.
For the values marked with $^{\dagger}$ (columns $B{=}2100$K), the MILP hit the 6000\,s timeout limit: 
for $p{=}2$, 3 of 5 workloads timed out; for $p{=}8$, all 5 did; 
the reported MILP incumbents are feasible but not guaranteed optimal. The second part of a cell shows the corresponding speedup achieved related to the sequential runtime.} 
\label{tab:pbm1_complex}
\end{table}

\begin{table}[t!]
\centering
\scriptsize
\begin{tabular}{|c|c|c|c|c|}
\hline
\multicolumn{5}{|c|}{\textbf{B = 210K}} \\ \hline
        & \multicolumn{2}{c|}{\textbf{X = 2}} 
        & \multicolumn{2}{c|}{\textbf{X = 4}} \\ \hline
\textbf{p} 
        & \textbf{MILP} & \textbf{GH} 
        & \textbf{MILP} & \textbf{GH} \\ \hline
\textbf{2} 
        & 99.81/1.61 & 99.60/1.84 
        & 100/1.9    & 96.84/1.86  \\ \hline
\textbf{4} 
        & 99.34/3.27 & 95.47/3.25 
        & 91.89/3.87 & 88.71/3.88  \\ \hline
\textbf{8} 
        & 100/4.26   & 98.83/3.82 
        & 66.61/4.33 & 92.21/7.35  \\ \hline
\end{tabular}

\vspace{4pt}

\begin{tabular}{|c|c|c|c|c|}
\hline
\multicolumn{5}{|c|}{\textbf{B = 630K}} \\ \hline
        & \multicolumn{2}{c|}{\textbf{X = 2}} 
        & \multicolumn{2}{c|}{\textbf{X = 4}} \\ \hline
\textbf{p} 
        & \textbf{MILP} & \textbf{GH} 
        & \textbf{MILP} & \textbf{GH} \\ \hline
\textbf{2} 
        & 95.24/1.98 & 94.84/1.99 
        & 54.83/1.94 & 74.74/1.99  \\ \hline
\textbf{4} 
        & 77.9/3.39  & 91.9/3.89  
        & 14.67/1.98 & 75.42/3.96  \\ \hline
\textbf{8} 
        & 15.27/1.41 & 81.24/7.44 
        & 2.94/1     & 74.03/7.9   \\ \hline
\end{tabular}

\caption{PBC - Heterogeneous transactions: MILP incumbent and GH rewards - $\sum_{\tau}\reward{\tau}$ of the block generated as a percentage of the LP-relaxation upper bound on achievable reward reported by the solver.}
\label{tab:pbm2_complex}
\end{table}

\subsection{Parallel-Block Construction}

For PBC on homogeneous workloads, we compare runtimes while varying the pool factor $X \in \{2,4,8\}$ and cores $p \in \{2,4,8\}$. 
MILP time again increases steeply with $X$ and $p$, while GH and RG remain in the millisecond range (\textbf{H1,H2}). 
For homogeneous mempool instances, GH and RG achieve reward values that are consistently close to the MILP optimum across all tested configurations ($X \in \{2,4,8\}$, $p \in \{2,4,8\}$, and $R \in \{10,30,100\}$). 
This shows that when execution times are uniform, greedy conflict-aware packing is sufficient to recover near-optimal block reward, supporting \textbf{H3}. 
Moreover, RG’s performance closely matches GH across the entire configuration grid, indicating that the simple reward-greedy baseline remains competitive on homogeneous workloads. Also, in this homogeneous setting both GH and RG achieve essentially perfect parallel utilization, yielding a speedup equal to the number of cores (i.e., exactly $p$) across the tested configurations, supporting \textbf{H4}.

For heterogeneous workloads, where both execution times and rewards vary, exact MILP optimization becomes prohibitive.  We therefore normalize both solver and heuristic rewards by the MILP’s LP-relaxation upper bound\footnote{The MILP solver used initially reports the LP-relaxation bound, obtained by relaxing integrality.}.  
\Cref{tab:pbm2_complex} presents these normalized values for pool factors $X \in \{2,4\}$, runtime limits $B \in \{210\text{K},630\text{K}\}$, and cores $p \in \{2,4,8\}$.  
GH schedules achieve between $74$–$100\%$ of the relaxation bound, while MILP incumbents frequently remain well below $20\%$ of the LP bound at large $B$ or $p$, having reached the time limit before closing the optimality gap (\textbf{H1,H2}). 
Even in the most challenging settings, GH constructs high-reward, feasible blocks within milliseconds, thereby confirming its scalability and practical robustness (\textbf{H3}).  
We also evaluated the reward-greedy baseline (RG) on the same instances and observed near-identical results to GH in this setting; for clarity, we therefore omit RG from the table. The speedup values obtained are close to the value of $p$, especially for large $B$ (\textbf{H4}), which means that using parallelism in this problem achieves almost linear speedup. 
Comparing with Table~\ref{tab:pbm1_complex}, we observe a big improvement when transaction selection is possible versus only scheduling a pre-determined block.

We conducted additional conflict-stress experiments with artificially high contention and found that GH is consistently at least as fast as RG and yields higher total reward while scheduling the same number of transactions, with an average improvement of 7.7\% over RG. In fact, a simple scenario in which GH will drastically improve over RG is one in which all transactions have execution time and reward 1, and the first $B$ transactions in the mempool conflict with each other, while all the other transactions (at least $p \cdot B$) conflict with the first $B$ but not with themselves. (This scenario may arise if the first $B$ transactions write in the same item $x$, while the rest reads $x$.) Then, RG will schedule the first $B$ transaction in sequence, not able to schedule other transactions in parallel. On the other hand, GH will schedule $p \cdot B$ transactions, always executing $p$ transactions in parallel.

\section{Conclusions} \label{sec:conclusion}

In this paper, we presented a systematic exploration of validator-side parallelism in blockchain transaction execution. We formulated two complementary optimization problems: (i) executing an already ordered block on multiple cores while minimizing makespan, and (ii) selecting and scheduling a subset of mempool transactions under a gas budget to maximize validator reward. For both, we developed exact MILP formulations to serve as optimal baselines, and efficient deterministic heuristics that scale to realistic workloads. We also include a Sealevel-inspired declared-access baseline (Sol) to represent a practical, conflict-avoiding in-order execution strategy.

Our evaluation on Ethereum mainnet traces demonstrate that MILP runtimes grow steeply with problem size, heterogeneity, and core count, confirming the computational hardness of the problems. Our heuristics remain sub-second across all settings, produce solutions that are close to those of the MILP, and reduce makespan and increase reward by a sensible amount with respect to sequential scheduling. Across our experiments, GH consistently matches or improves upon Sol in solution quality, with particularly clear advantages as conflict intensity increases, while maintaining comparable (sub-second) scheduling overhead.

From a systems perspective, our findings show that even simple, conflict-aware scheduling strategies can make effective use of available cores, narrowing the gap between theoretical and practical parallelism in blockchain execution. In particular, the mempool-based formulation provides a path toward constructing blocks that are inherently “parallel-friendly,” coupling selection and scheduling to improve both performance and economic efficiency.

\textbf{Limitations and Future Work.} Our evaluation uses the \texttt{gasUsed} field from Ethereum traces as a proxy for execution time. This provides a consistent and realistic measure for the ordered-block problem, but it is only an approximation of the actual runtime. In practice, execution time and gas usage may diverge due to hardware, client, or EVM implementation differences. Moreover, in the mempool formulation, reordering transactions could change their effective gas consumption in the case of conflicts. A second limitation is that our current heuristics largely resolve conflicts through local, greedy decisions based on static conflict information. A promising direction is to design more holistic conflict-aware policies that reason globally about contention (e.g., hotspots, conflict clusters, or anticipated serialization) when selecting and scheduling transactions.

Exploring how to model such non-determinism in execution time while maintaining the safety guarantees provided by conservative gas limits is left for future work. Similarly, an open systems question is how best to maintain and update dependency graphs efficiently as new transactions arrive and blocks are mined. This also suggests online heuristics that update priorities as conflicts evolve, rather than relying on a fixed, one-shot ordering. Designing low-overhead data structures or incremental graph-maintenance schemes for this purpose will be essential for bringing these scheduling techniques into production blockchain environments.

\bibliographystyle{plainurl}
\bibliography{sample}




\end{document}